\documentclass[aps,prl,twocolumn,showpacs,floatfix]{revtex4}
\usepackage{epsfig}
\usepackage{psfrag}
\usepackage{graphicx}
\begin{document}
\title{Fragmentation of shells}

\author{ F.\ Wittel${}^1$, F.\ Kun${}^{2}\footnote{Electronic 
address:feri@dtp.atomki.hu}$, H.\ J.\ Herrmann${}^3$, and B.\ H.\
Kr\"oplin${}^1$} 
\affiliation{
${}^1$Institute of Statics and Dynamics of Aerospace Structures,
University of Stuttgart, Pfaffenwaldring 27, 70569 Stuttgart,
Germany\\  
${}^2$Department of Theoretical Physics,
University of Debrecen, P.\ O.\ Box:5, H-4010 Debrecen, Hungary \\
${}^3$ICA 1, University of Stuttgart, Pfaffenwaldring 27, D-70569
Stuttgart, Germany} 
\date{\today}
   
\begin{abstract}
We present a theoretical and experimental study of the fragmentation 
of closed thin shells made of a disordered brittle material.
Experiments were performed on brown and white hen egg-shells
under two different loading conditions: 
impact with a hard wall and explosion by a combustible mixture 
both give rise to power law fragment size distributions.
A three-dimensional discrete element model of shells is
worked out. Based on simulations of the model we give evidence
that power law fragment mass distributions arise due to an underlying 
phase transition which proved to be abrupt for explosion and continuous
for impact. We demonstrate
that the fragmentation of closed shells defines a new universality
class of fragmentation phenomena.
\end{abstract}
 

\maketitle

Closed shells made of solid materials are often used in every day
life and industrial applications in form of containers,
pressure vessels or combustion chambers. From a structural point of view
aircraft vehicles, launch vehicles like rockets and building blocks of
a space station are also shell-like systems and even
certain types of modern buildings can be considered as shells. The
egg-shell as nature's oldest container proved to be a reliable
construction for protecting life. 
In many of the applications shell-like constructions operate under an
internal pressure and they usually fail due to an excess internal load
which can arise either as a result of slowly driving the system above
its stability limit, or by a pressure
pulse caused by an explosive shock inside the shell. Due to the
widespread applications, the failure of shell systems is a very
important scientific and technological problem which can also have  
an enormous social impact due to the human costs arising, for
instance, in accidents.

Fragmentation, {\it i.e.} the breaking of particulate materials into
smaller pieces is abundant in nature occurring on a broad range of
length scales from meteor impacts through
geological phenomena and industrial applications 
down to the break-up of large molecules and heavy nuclei \cite{exp_turc,exp_bohr,kadono,exp_inao,katsuragi}.
The most striking observation concerning fragmentation is that the
distribution of fragment sizes shows a power 
law behavior independent on the way of imparting energy, relevant
microscopic interactions and length scales involved, with an exponent
depending only on the dimensionality of the system
\cite{exp_turc,exp_bohr,kadono,exp_inao,katsuragi,ferenc1,astrom1,astrom2,ching}.
Detailed experimental and theoretical studies
revealed that universality prevails for large enough input energies
when the system falls apart into small enough pieces
\cite{exp_turc,exp_bohr,katsuragi,kadono,exp_inao}, however, at 
lower energies a systematic dependence of the exponent on the input
energy is evident \cite{ching}. Recent investigations on the low
energy limit of fragmentation suggest that the power law distribution
of fragment sizes arises due to an underlying critical point
\cite{katsuragi,ferenc1,astrom1,astrom2}. 
Former studies on
fragmentation have been focused on bulk systems in
one, two and three dimensions, however, hardly 
any studies have been devoted to the fragmentation of shells. The
peculiarity of the fragmentation of closed shells originates from the
fact that their local structure is inherently two-dimensional, however,
the dynamics of the systems, the motion of material elements,
deformation and stress states are three-dimensional which allows for a
rich variety of failure modes.

 \begin{figure}
\begin{center}
\epsfig{bbllx=20,bblly=20,bburx=575,bbury=160,file=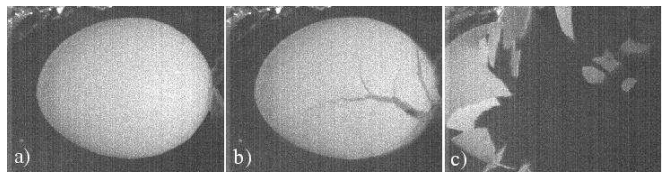,
  width=8.7cm}
 \caption{\small Time evolution of the explosion of an egg-shell.
}
\label{fig:eggsplode}
\end{center}
\end{figure} 
In this letter we present a theoretical and experimental 
study of the fragmentation of closed brittle shells arising due to an
excess load inside the shell. We performed experiments on the
explosion and impact fragmentation of hen egg-shells resulting in power
law fragment size distributions. 
Based on large scale molecular dynamics simulations of a discrete
element model of shells we give
evidence that power law fragment mass distributions arise due to an
underlying phase transition which proved to be abrupt for explosion
and continuous for impact.
It is shown that the fragmentation of closed shells belongs to a
universality class different from that of the two- and
three-dimensional bulk systems.
Our results have also important implications for the fragmentation of
bulk systems.

For simplicity, our experiments were performed on ordinary brown and
white egg-shells  
which consists of a natural disordered material with 
a high degree of brittleness. In the preparation, first two holes of
regular circular shape were drilled in the bottom and top of the egg
through which the content of the egg was blown-out. The inside
was carefully washed and rinsed out several times and finally the empty
shells were dried in a microwave oven to get rid of all moisture of the
egg-shell. 

In the impact experiments, intact egg shells are catapulted against the
ground at a high speed using a simple setup of rubber bands. 
In explosion experiments, initially the egg shell is flooded with
hydrogen and  
hung vertically inside a plastic bag. The combustion reaction is
initiated by igniting the 
escaping hydrogen on the top of the egg. 
Due to the escaping hydrogen, oxygen is drawn-up into the egg through
the bottom hole.
When enough air has entered to form a combustible
mixture inside the egg, the flame penetrates through the top hole and
the egg explodes. Both kinds of experiments were
carried out inside a soft plastic bag so that secondary fragmentation
due to the impact of fragments did not occur.

The resulting egg-shell pieces are then carefully collected and placed
on the tray of a scanner without overlap. In the scanned image
fragments appear as black spots on a white background,
which were further analyzed by a cluster searching code. The mass of
fragments was determined as the number of pixels of spots in the
scanned image. 

\begin{figure}
\begin{center}
\epsfig{bbllx=20,bblly=20,bburx=570,bbury=480,file=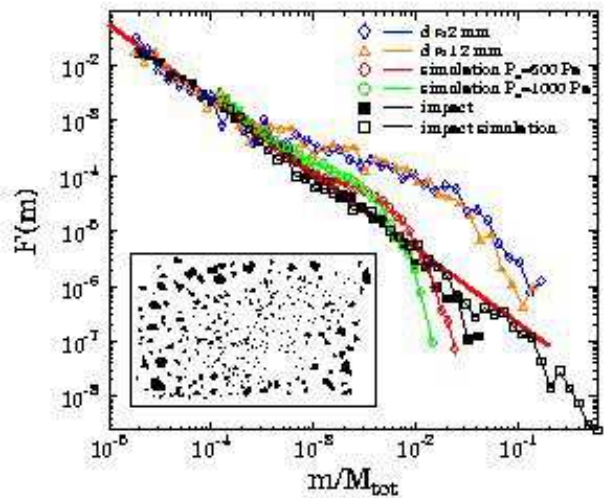,
      width=8.2cm}
 \caption{\small {\it (Color online)} Comparison of fragment mass
   distributions obtained in explosion experiments with two hole
   sizes, and in the impact experiment to the simulation
   results. Inset: scanned pieces of an impact experiment.
}
\label{fig:exp_imp}
\end{center}
\end{figure}

Three consecutive snapshots of the explosion
process of an egg-shell are presented in Fig.\ \ref{fig:eggsplode}
taken by a high speed camera of 400 Hz.
Based on the
snapshots the total duration of an explosion is estimated to be of the
order of 1 msec. An example of scanned egg-shell pieces
can be seen in the inset of Fig.\ \ref{fig:exp_imp}.
Since the pressure which builds up during combustion can be
slightly controlled by the hole size, {\it i.e.} the smaller the hole
is, the higher the pressure at the explosion is, we performed several
series of explosion experiments with hole diameters $d$ between 1.2
and 2.5 millimeters. Fig.\ \ref{fig:exp_imp}
presents the fragment mass distributions $F(m)$ for impact
and explosion experiments averaged over 10-20 egg-shells for each
curve. For the impact experiment, a power law behavior of
the fragment mass distribution $ F(m) \sim m^{-\tau}$
can be observed over three orders of magnitude where
the value of the exponent can be determined with high precision
to $\tau = 1.35 \pm 0.02$. Explosion 
experiments result also in a power law distribution of the same value
of $\tau$ for small fragments with a 
relatively broad cut-off for the large ones. 
Smaller hole diameter $d$ in Fig.\ \ref{fig:exp_imp}, {\it i.e.}
higher pressure, gives rise to a  
larger number of fragments with a smaller cut-off mass and a faster
decay of the distribution $F(m)$ at the large fragments. Comparing the
number of fragments obtained, the ratio of the pressure values 
in the explosions at hole diameters $d=1.2$ and 2.0 mm, presented
in Fig.\ \ref{fig:exp_imp}, was estimated to be about 1.6.
Note that the relatively
small value of the exponent $\tau$ can indicate a cleavage mechanism
of shell fragmentation and is significantly different from the
experimental and theoretical results on fragmenting two-dimensional
bulk systems where $1.5 \leq \tau \leq 2$ has been found, and from the
three-dimensional ones where $\tau > 2$ is obtained 
\cite{exp_turc,exp_bohr,kadono,exp_inao,katsuragi,ferenc1,astrom1,astrom2,ching}.

Our theoretical study is restricted for simplicity to spherical
shells, {\it i.e.} we worked out a three-dimensional discrete element
model of spherical shells by discretizing the surface of the unit
sphere into randomly shaped triangles (Delaunay triangulation).
The nodes of the triangulation represent point-like material  
elements in the model whose masses are defined by the area of the dual
Voronoi polygon \cite{lauritsen} assigned to it. The bonds between
nodes are assumed to be springs having linear elastic behavior up to
failure.  
Disorder is introduced in the model solely by the randomness of the
tessellation so that the masses of the nodes, the
lengths and the cross-sections of the springs are determined by the
tessellation (quenched structural disorder). After prescribing the
initial conditions of a specific fragmentation process studied, the
time evolution of the system is 
followed by solving the equation of motion of the nodes.
In order to account for crack formation in the model, springs are
assumed to break during the time evolution of the system when their
deformation $\varepsilon$ exceeds a fixed breaking 
threshold $\varepsilon_c$ resulting in a random sequence of
breakings due to the disordered spring properties. 
As a result of
successive spring breakings cracks nucleate, grow and merge on the
spherical surface giving rise to the complete break-up of the
shell into pieces. Fragments of the shell are defined in the
model as sets of nodes connected by the remaining intact springs.   
The process is stopped when the system has attained a relaxed state.

\begin{figure}
\begin{center}
\epsfig{bbllx=20,bblly=20,bburx=575,bbury=280,file=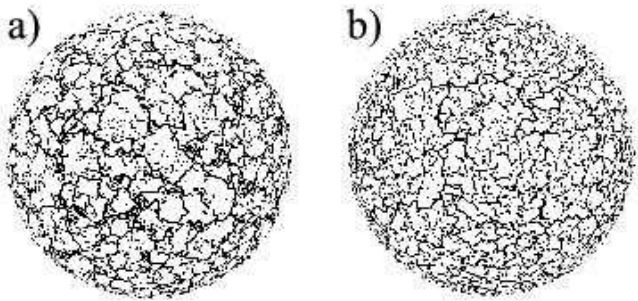,
  width=8.5cm}
 \caption{\small 
Final fragmented states for $a)$ impact and $b)$ pressure
pulse loading. Particle positions are projected back to the initial
state on the surface. Black lines indicate cracks.}
\label{fig:cracks}
\end{center}
\end{figure}
In the computer simulations two different
ways of loading have been considered which model the experimental
conditions and represent limiting cases of energy input rates:
(i) {\it pressure pulse} and (ii) {\it impact} load starting from an
initially stress free state. A pressure pulse
in a shell is carried out by imposing a fixed internal pressure $P_o$ 
from which the forces acting on the material elements are calculated. 
The constant pressure $P_o$ gives rise to an expansion of the
system with a continuous increase of the imparted energy $E_{tot} =
P_o\Delta V$, where $\Delta V$ denotes the volume change with respect
to the initial volume $V_o$. 
Since the force $F$ acting on the shell is
proportional to the actual surface area $A$, the system is driven
by an increasing force $F \sim P_o \cdot A$ during the expansion
process.
The impact loading realizes the limiting case of instantaneous energy
input $E_{tot}=E_o$ by giving a fixed initial radially oriented
velocity $v_o$ to all material elements. 
Simulations performed varying
the control parameters $P_o$ and $E_o$ in a broad range revealed that
for both loading cases there exists a critical value $P_c$ and $E_c$
below which the shell keeps its integrity suffering only partial
failure in the form of cracks (damaged state), above it however,
complete break-up occurs into pieces (fragmented state). Since the
break-up of the shell under pressure (impact) loading is analogous to
the stress (strain) controlled fracture of disordered bulk solids, the
transitions between the damaged and fragmented states arises abruptly
(continuously) at the critical point.
\begin{figure}
\begin{center}
\epsfig{bbllx=20,bblly=20,bburx=570,bbury=480,file=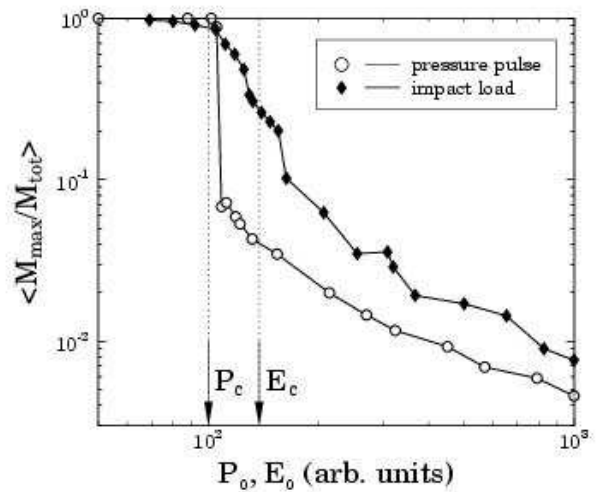,
  width=8.0cm}
 \caption{\small $\left<M_{max}/M_{tot}\right>$ versus
   $P_o$ and $E_o$. The values of $P_c$ and $E_c$ are also indicated.
}
\label{fig:maxmass}
\end{center}
\end{figure}

\begin{figure}
\begin{center}
\epsfig{bbllx=20,bblly=20,bburx=570,bbury=480,file=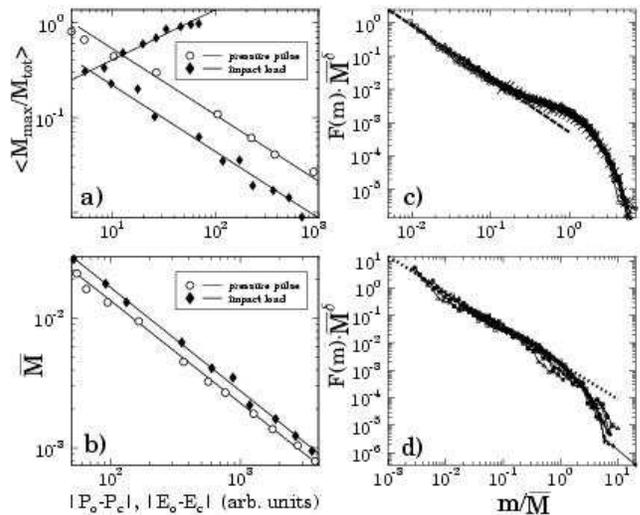,
  width=8.6cm}
 \caption{\small Normalized mass of the largest fragment $a)$ and
   average fragment mass $b)$ as a function
   of the distance from the critical point.
   Rescaled plots of the fragment mass distributions $c)$ for explosion,
   and $d)$ for impact.
}
\label{fig:massdist}
\end{center}
\end{figure}

Quantitative characterization of the break-up process
when increasing the control parameter
can be given by monitoring the average mass of the largest fragment
normalized by the total mass $\left<M_{max}/M_{tot}\right>$.
$M_{max}$ is a monotonically decreasing function of both $P_o$ and
$E_o$, however, the functional forms are different in the two cases,
see Fig.\ \ref{fig:maxmass}. 
At low pressure values in Fig.\ \ref{fig:maxmass} $M_{max}$ is
practically equal to the total mass since 
 hardly any fragments are formed.
Above $P_c$ however, $M_{max}$ gets
 significantly smaller than $M_{tot}$ indicating the disintegration of
 the shell into pieces. 
 The value of the critical pressure $P_c$ needed to achieve
fragmentation and the functional form of the curve  above $P_c$ were
determined by plotting $\left<M_{max}/M_{tot}\right>$  as a
function of the difference $|P_o-P_c|$ varying $P_c$ 
until a straight line is obtained in a double logarithmic plot. 
The power law dependence on the distance from the critical point $
\left<M_{max}/M_{tot}\right> \sim |P_o - P_c|^{-\alpha}$
for $P_o > P_c$ is evidenced by Fig.\ \ref{fig:massdist}$a)$, where
$\alpha = 0.66 \pm 0.02$ was determined.  The finite jump of $M_{max}$
at $P_c$ indicates the abrupt nature of the transition between the two
regimes.
For impact loading $\left<M_{max}/M_{tot}\right>$ proved to be a
continuous function of $E_o$, however, it also shows the existence of
two regimes with a transition at a critical
energy $E_c$. In Fig.\ \ref{fig:massdist}$a)$ we show
$\left<M_{max}/M_{tot}\right>$ as a function of the distance
from the critical point $|E_o - E_c|$, where $E_c$ was determined
numerically in the same way as $P_c$. As opposed 
to the pressure loading, $\left<M_{max}/M_{tot}\right>$ exhibits a
power law behavior on both sides of the critical point but with
different exponents
$   \left<M_{max}/M_{tot}\right> \sim |E_o - E_c|^{\beta}$
for $E_o < E_c$ and $\left<M_{max}/M_{tot}\right> \sim |E_o -
E_c|^{-\alpha}$ for $E_o > E_c,$ where the exponents were obtained as
$\beta = 0.5 \pm 0.02$ and $\alpha = 0.66 \pm 0.02$.
Note that the value of $\alpha$ coincides with the
corresponding exponent of the pressure loading.

We also evaluated the mass-weighted average fragment mass $\overline{M}$ as
the average value of the ratio of the second $M_2$ and first $M_1$
moments of the fragment mass distribution subtracting the mass of the
largest fragment from the sum \cite{ferenc1}.
The behavior of $\overline{M}$ shows again clearly
the existence of two regimes of the break-up process with a transition
at the critical point $P_c$ and $E_c$.
Under pressure loading due to the abrupt disintegration $\overline{M}$
can only be evaluated above the critical point $P_c$, while for the
impact case $\overline{M}$ has a maximum at the critical energy $E_c$
typical for continuous phase transitions.  Fig.\ \ref{fig:massdist}$b)$
demonstrates that in both loading cases
$\overline{M}$ has a power law dependence on the distance from the
critical point, {\it i.e.} $\overline{M} \sim |E_o-E_c|^{-\gamma}$, and
 $\overline{M} \sim |P_o-P_c|^{-\gamma}$ with the same value of the
 exponent $\gamma = 0.8\pm0.02$.

The most important characteristic quantity of our system which can
also be compared to the experimental results is the mass distribution
of fragments $F(m)$. For
pressure loading $F(m)$ can only be evaluated above $P_c$.
Under impact loading for $E_o < E_c$ we found that $F(m)$
has a pronounced peak at large fragments indicating the presence of
large damaged pieces. Approaching 
the critical point $E_c$ the peak gradually disappears and the
distribution asymptotically becomes a power law at $E_c$. Above the
critical point 
the power law remains for small fragments followed by an exponential
cut-off for the large ones. For the purpose of comparison, a mass
distribution $F(m)$ obtained at an impact energy close to the critical
point $E_c$, and distributions at two different pressure values $P_o$
of the ratio 1.6 are plotted in Fig.\ \ref{fig:exp_imp} along with the
experimental results. For impact an excellent agreement of the
experimental and theoretical results is evidenced. For pressure
loading, the functional form of $F(m)$ has a nice qualitative
agreement with the experimental findings on the explosion of eggs,
furthermore, distributions at the same ratio of pressure
values obtained by simulations and experiments proved to show the
same tendency of evolution, see Fig.\ \ref{fig:exp_imp}. 
Figs.\ $\ref{fig:massdist}c,d)$ demonstrate that by
rescaling the mass distributions above the critical point by plotting
$F(m) \cdot \overline{M}^{\delta} $ as a function of $m/\overline{M}$ an
excellent data collapse is obtained with $\delta = 1.6\pm 0.03$. The
data collapse implies the validity of the scaling form $F(m) \sim
m^{-\tau} \cdot f(m/\overline{M})$, typical for critical
phenomena. The cut-off function $f$ has a simple exponential form
$exp(-m/\overline{M})$ for impact loading, and a more complex one
containing also an exponential component for the pressure case.  
The rescaled plots make possible an accurate determination of the
exponent $\tau$, where $\tau = 1.35\pm 0.03$ and $\tau = 1.55\pm 0.03$
were obtained for impact and pressure loading, respectively. 
Hence, a good quantitative agreement of the theoretical and experimental
values of the exponent $\tau$ is
evidenced for the impact loading of shells, however, for the case of
pressure loading the numerically obtained exponent turned out to be
higher than in the case of exploded eggs. 

In the fragmentation of bulk systems under appropriate conditions
a so-called detachment effect is observed when a
surface layer breaks off from the bulk and undergoes a separated
fragmentation process \cite{ferenc1,ching}. This
effect also shows up 
in the fragment mass distributions in the form of a power law regime
of small fragments of an exponent smaller than for the large ones.
Our results on shell fragmentation can also provide a
possible explanation of this kind of composite power laws of bulk
fragmentation \cite{ferenc1,ching}, and 
might even be relevant for the study of mass redistribution following
super nova explosions in the universe.  

\begin{acknowledgments}
This work was supported by the projects SFB381 and OTKA T037212,
M041537. F.\ Kun was supported by the FKFP 0118/2001 and by the
Gy. B\'ek\'esi Foundation of HAS.  
\end{acknowledgments}

\end{document}